# Implications of Russia's full-scale invasion of Ukraine for the international mobility of Ukrainian scholars

**Corresponding author**
Myroslava Hladchenko
Center for R&D Monitoring, University of Antwerp, Belgium
hladchenkom@gmail.com, myroslava.hladchenko@uantwerp.be

**Abstract**
This study explores the implications of Russia's full-scale invasion of Ukraine for the international mobility of Ukrainian scholars. The dataset, drawn from the CWTS in-house Scopus database, comprises Ukrainian scholars who were internationally mobile between 2020 and 2023. The analysis focuses on scholars affiliated with universities and the National Academy of Sciences of Ukraine (NASU) before moving abroad. The findings revealed a rise in the number of internationally mobile scholars in 2022–2023, driven primarily by increased mobility from universities. For NASU, Russia was the top destination country in 2020–2021 but dropped to fourth place in 2022–2023, overtaken by Germany, China, and Poland. For universities, Poland, Germany, and Russia consistently ranked as the top three destination countries across both periods. A statistical test indicated no significant difference in mean FNCI between scholars who were internationally mobile in 2020–2021 and 2022–2023. However, the share of internationally mobile scholars with articles among the top 10% most cited globally increased among those previously affiliated with universities but declined among those affiliated with NASU. In both periods, the share of scholars with articles in the top 10% most cited globally, published during the five years before changing the country of affiliation, was higher among internationally mobile scholars than among those with Ukrainian affiliations. Whether this mobility implies a brain drain requires further research. If effectively leveraged, it may strengthen Ukraine's integration into global scientific networks, support post-war recovery, and foster a more resilient, internationally connected, and competitive academic system.
**Keywords**: Russian invasion of Ukraine, international mobility, destination countries for mobile scholars, citation impact.

## Introduction

International mobility is widely studied for its role in facilitating research collaboration, knowledge flows, and scholarly careers (Horta et al., 2010; Stephan, 2012; Robinson-Garcia et al., 2019; Shen et al., 2022). International mobility also boosts collaboration (Bologna Follow-Up Group, 2009), personal development and research excellence (European Commission, 2012; Sautier, 2021). It broadens scientists' views on issues affecting the whole world (Rodrigues et al., 2016). International mobility can be divided into short-term research visits and academic migration when scholars decide to remain in another country (Lauder, 2005). However, most of the prior studies have focused on voluntary mobility driven by career opportunities, institutional attractiveness, or personal development. Far less attention has been paid to mobility under conditions of coercion—when war forces scholars to migrate. In recent decades, cases of displaced scholars have been reported in contexts such as Afghanistan, Syria, and, most recently, Ukraine (Beaney, 2024; De Carvalho et al., 2023; Tarkhanova & Pyrogova, 2024; Akkard, 2025; Chala et al., 2024; Oleksiyenko et al., 2023). International mobility under duress differs fundamentally from voluntary mobility, as it raises pressing questions regarding the continuity of research, the rebuilding of higher education, and the dynamics of global knowledge flows during crises.

Ukraine provides a particularly important case for studying forced academic mobility. As a post-Soviet country, its transition was shaped by the capture of governmental institutions by Soviet-era elites and actors from the shadow economy (Riabchuk, 2009; Kudeila, 2012). Oligarchic dominance fostered rent-seeking and blocked radical reforms (Hellman, 1998), leaving

Ukraine with a low-tech economy dependent on raw materials—80% of exports were semi-finished products (Yegorov & Ranga, 2014). The lack of a coherent strategy for reforming industry, higher education, and science contributed to early waves of scholar migration in the 1990s and rendered academic careers unattractive to younger generations in the following years (Hladchenko, 2023).

In 2013, Ukrainians started a Revolution of Dignity because the pro-Russian president refused to sign an Association Agreement with the EU (Umland et al., 2014; Umland, 2016). Ukrainians rebelled against the exploitation of governmental institutions, which implied a high level of corruption and lawlessness. Although the revolution succeeded, Russia responded with invasion in 2014 (Kazdobina et al., 2024; Kuzio, 2022). This invasion caused the displacement of scholars and universities within the country (Oleksiyenko et al., 2021). The full-scale invasion in 2022 intensified these dynamics, leading to mass displacement of scholars within and outside Ukraine. More than one out of ten Ukrainian scholars (12%), namely 10,429 individuals, were forced either to emigrate abroad (5,542; 6.3%) or to become internally displaced persons (4,887; 5.5%). In addition, 1,518 scholars joined the army to defend Ukraine (Cabinet of Ministers of Ukraine, 2024). A total of 146 Ukrainian scholars were killed, comprising individuals who served in the army as well as those killed in Russian missile attacks (Novynarnia, 2024). Russia's full-scale invasion also prompted international responses from international community (Nazarovets & Teixeira da Silva, 2022), including governments, universities, and funding bodies aimed at providing Ukrainian scholars with fellowships for temporary stays abroad, including initiatives by the US National Academies of Sciences, the European Commission, and the Aleksander von Humboldt Foundation (OECD, 2022).

This study investigates how Russia's full-scale invasion of Ukraine has affected the international mobility of Ukrainian scholars, focusing on scale of mobility, top destination countries, disciplinary shifts, and citation impact. The analysis examines scholars affiliated with universities and the National Academy of Sciences of Ukraine, which together employ the majority of Ukrainian scholars. Using Scopus data, the study compares two years before (2020–2021) with two years during Russia's full-scale invasion (2022–2023).

**International mobility**
Recent science policies emphasise academic mobility as a mechanism for fostering international networks and enhancing competitiveness (EC, 2010; OECD, 2008). This policy orientation reflects a belief that mobility promotes knowledge recombination and better matches between scholars and institutions, whereas immobility may hinder productivity (Horta et al., 2010; Stephan, 2012).

Earlier debates often framed international mobility as beneficial only to host countries and detrimental to countries of origin (Cañibano et al., 2016). However, more recent studies demonstrate that mobility benefits both sending and receiving countries by facilitating knowledge flows (Robinson-Garcia et al., 2019; Shen et al., 2022; Saxenian, 2002). Mobile scholars help sustain links between home and host countries, enabling the transfer of knowledge back to the origin country (Jonkers & Cruz-Castro, 2013; Kahn & MacGarvie, 2016). They also act as bridges between non-mobile colleagues and global research networks (Franzoni et al., 2012; Fry, 2022). Such linkages positively affect scientific productivity in both host and home systems (Baruffaldi & Landoni, 2012).

Mobility also provides scholars with access to resources not available in their home countries, including funding, advanced equipment, and specialised expertise (Netz et al., 2020). Recognising these benefits, many public agencies actively promote academic mobility. For example, the Swiss National Science Foundation, the German Academic Exchange Service, and the Polish National Agency for Academic Exchange fund international study visits to foster collaboration and knowledge exchange.

Individual motivations are equally significant. According to the GlobSci survey, scholars most often cite career opportunities and access to outstanding research teams as primary reasons for moving abroad. Other pull factors include dynamic and competitive funding systems (Van Noorden, 2012), better working conditions, and a higher quality of life (Laudel, 2005). Conversely, push factors include limited funding, lack of attractive research positions, obsolete infrastructure, and weak labour markets in home countries (Ackers, 2005; Laudel, 2005).

Disciplinary characteristics shape the extent of international mobility. The physical sciences and engineering are marked by high levels of geographical mobility due to large-scale international projects and specialised facilities concentrated in certain locations (Petersen, 2018; Hunter et al., 2009; Sincell, 2000; Stephan, 2012; Scellato et al., 2012). Life sciences also display high mobility, whereas the social sciences are the least mobile (Momeni et al., 2022; Laudel & Bielick, 2019).

The richer a country is, the more scholars flock to it (Van Noorden, 2012). The USA and China are two large hubs, and England and Germany are two smaller hubs for highly mobile scholars (Aref et al., 2019). The USA, Germany and the UK are among the rich economies with the largest number of migrants (Grogger & Hanson, 2011). The USA has always been a hub for highly skilled migrants, and it has been considered a top destination for advancing a research career (Van Noorden, 2012; Franzoni et al., 2012; Sincel, 2000). National policies define the flow of scholars to a country. In 2008, China initiated a 'One Thousand Talents Scheme' to recruit academics from abroad. In 2024, China announced possible changes to the immigration system to attract more foreign scientists, including offering permanent residence (Bela & Peng, 2024). In 2001, Germany's Federal Ministry of Education and Research (BMBF) announced a scheme to produce a 'brain gain' by acquiring foreign and German scientists working abroad (BMBF, 2001). Germany maintains international mobility through DAAD scholarships and the USA through the Fulbright fellowship.

Cultural and linguistic proximity also influences migration choices, as shared language and cultural ties reduce integration barriers (Ackers, 2005; Adsera & Pytlikova, 2015). Additionally, established scientific collaborations guide the choice of host country (Appelt et al., 2015). While salary can motivate academics to move abroad, it is generally not the primary factor (Ackers, 2005; Richardson & Mallon, 2005; Austin et al., 2014).

The effects of international mobility depend on multiple factors, including characteristics of the host and home countries, gender, discipline, institutional prestige, and duration of mobility (Gu et al., 2024; Singh, 2018; Finocchi et al., 2023). Mobile scholars are generally more productive in terms of publication output than their non-mobile colleagues, though the effect can be marginal (Ganguli, 2011; Aksens et al., 2013). Conversely, highly productive scholars are also more likely to engage in international mobility (Franzoni et al., 2012; Hunter et al., 2009). Beyond productivity, internationally mobile scholars tend to produce more impactful publications (Finocchi et al., 2023; Dubois et al., 2014; Franzoni et al., 2014; OECD, 2015). Citation counts for mobile scholars can be up to 17% higher than for non-mobile colleagues, largely due to increased diversity of co-authors and research topics after migration (Petersen, 2018; Gu et al., 2024). Overall, the scientific impact of mobile scholars is around 20% higher than that of non-mobile scholars (OECD, 2013), and they also publish in more prestigious journals (Aykac, 2021)

**War and international mobility**
In recent decades, war has increasingly become a driver of scientific mobility, displacing scholars from countries such as Afghanistan, Syria, and, most recently, Ukraine (Beaney, 2024; De Carvalho et al., 2023). The term "displaced academics," first used in the context of forced migration in Nazi-occupied Europe, was revived following the Russian annexation of Crimea and the invasion of Donbas in 2014 (Oleksiyenko et al., 2021). It refers to scholars who are compelled to leave their home country due to war, conflict, or political instability. Providing support to displaced academics enables them to continue their research through access to international networks and resources (De Carvalho et al., 2023).

As a rule, developed countries host displaced scholars. Germany is a particularly attractive destination, given its extensive third-party funding opportunities. At the same time, however, the German academic system faces a chronic shortage of permanent positions, limiting long-term prospects for displaced researchers (Vatansever, 2022). Similar patterns—abundant short-term fellowships but scarce permanent posts—are reported in the UK (Beaney, 2024; Akkad, 2022).

The choice of host country often depends on political, social, and academic networks, as well as on national or organisational funding initiatives (Yarar & Karakaşoğlu, 2022). For example, many Ukrainian scholars who fled to Poland after the 2022 full-scale invasion highlighted geographic proximity, which allowed them, often as women with children, to maintain contact with family members who remained in Ukraine. Prior academic ties with Polish colleagues were also an important factor (Kiselyova & Ivashchenko, 2024). In contrast, Ukrainian scholars who moved to Brazil were motivated not only by the search for physical safety but also by curiosity about Brazilian culture and a desire for a multicultural experience (De Carvalho et al., 2023).

Hosting displaced academics is a mutually beneficial arrangement (McGrath & Lempinen, 2021). For host countries, displaced scholars represent a "brain gain" (Fierros-Pesqueira & Castillo-Federico, 2022; Oleksiyenko, 2021). Over time, their innovative, entrepreneurial, and scientific contributions can significantly enrich host societies—scientifically, socially, and culturally (Ergin et al., 2019).

**Overview of the system of higher education and science in Ukraine**

Since 1991, Ukraine has moved away from the Soviet political and economic order, but its system of higher education and science has remained deeply embedded in Soviet structures. Ukraine retained the institutional separation between predominantly teaching-oriented higher education institutions and the research institutes of the National Academy of Sciences of Ukraine (NASU). This differentiation contrasts with Western systems, where even in countries with non-university research institutes (e.g., Germany), universities still combine teaching and research functions (Clark, 1983).

NASU preserved its hierarchical governance structure and Soviet-style administrators, many of whom had longstanding ties to the communist party (Josephson & Egorov, 1997). The Presidium, a board of 34 academicians, directs the Academy. In 1991, 47,000 researchers were affiliated with NASU; by 2023, this number had fallen to 13,883 (NASU, 2023). The Academy's institutes are organised into fourteen disciplinary sections, ranging from physics and mathematics to economics, history, and law. Physics remains the dominant field, although the social sciences and humanities are represented by 31 institutes and two research libraries. NASU is primarily supported through block funding, a portion of which is redistributed as project funding by the Presidium.

In the early 1990s, most higher education institutions were reclassified as universities and formally authorised to conduct research (Parliament of Ukraine, 2002; 2014). However, the government neither established adequate research infrastructure nor provided dedicated research funding (Hladchenko et al., 2020). According to legislation, academics have a 36-hour working week, equivalent to 1,548 hours per year, divided among teaching, research, methodological work (e.g., writing textbooks), and administrative duties (Parliament of Ukraine, 2002). In 2014, the teaching load was reduced from 900 to 600 hours to expand time for research (Ministry of Education & Science of Ukraine, 2016). Nevertheless, in practice, academics remain overloaded with administrative and even non-academic tasks (Hladchenko & Westerheijden, 2021).

Universities do not receive basic research funding but may apply for competitive project funding from the Ministry of Education and Science. Since 2020, scholars affiliated with both universities and NASU have been eligible to apply to the National Research Foundation of Ukraine (NRFU), which distributes state funding as well as funds from foreign agencies for joint projects with international partners.

Both national (Scientific Committee of the National Board of Ukraine on Science and Technology, 2023; De Rassenfosse, Murovana, & Uhlbach, 2023; Bezvershenko & Kolezhuk, 2022) and international experts (Horizon 2020 Policy Support Facility, 2017; Schiermeier, 2019) argue that Ukraine must decisively break with these Soviet-style structures, cultures, and practices, which continue to hold back the development of higher education and science.

**Research assessment policies in Ukraine**

After 1991, research assessment policies in Ukraine considered articles only in Ukrainian journals. Thus, publications of Ukrainian scholars were mainly invisible to the European research community. The two-level systems of doctoral degrees and scientific titles (docent and professor) were also inherited unchanged from the Soviet period. In 2013, the education ministry introduced publications in international journals and Ukrainian Scopus- and WoS-indexed journals in research assessment policies, in particular, doctoral publication requirements, striving for Ukrainian universities to take higher positions in global rankings (Ministry of Higher Education and Science, Youth and Sport of Ukraine, 2012) (Table 1). In 2015, publications in Scopus- and WoS-indexed journals got a place in the licensing of study programmes requirements (Cabinet of Ministers of Ukraine, 2015). In the following years they were added to the requirements for the scientific titles of associate professor and professor (Ministry of Education and Science of Ukraine, 2016) and criteria for research assessment of higher education institutions (Cabinet Ministers of Ukraine, 2017).

**Table 1** Summary of requirements for (international) publications in Ukrainian research assessment policies

| Policy | Publication requirements |
| --- | --- |
| Doctoral degrees (2013) | Publications in Ukrainian Scopus- and WoS-indexed journals or in international journals<br>Since 2020 publications in any Scopus- and WoS-indexed journals<br>Since September 2021 for the Doctor of Sciences and since September 2020 for PhD, the number of articles needed for their obtention can be reduced by publishing in Q1-Q3 journals |
| Licensing (2015) | At least five articles published during last five years in Ukrainian journals, in journals indexed in databases including Scopus and WoS |
| Requirements for scientific titles of associate professor and professor (2016) | Article(s) in Scopus- and WoS indexed journals |
| Research assessment of higher education institutions (2017) | Articles in journals indexed in databases (the legislation does not clarify the names of databases) |
| Funding of research projects (2019) | Publications in Scopus- and WoS-indexed journals |
| Performance-based funding (2019) | Indicator of international recognition is defined based on the positions in international university rankings. |

According to the OECD (2022), Ukrainian scientific output demonstrates above-average specialisation and expertise, as measured by citation impact, particularly in computer science and energy. Ukrainian nuclear engineers are actively involved in nuclear programs worldwide. Between 2006 and 2020, the share of publications by Ukrainian scholars in the top 10% most-cited globally increased from 2% to 6%, reflecting an overall improvement in research quality by international standards (OECD, 2022).

However, previous studies have shown that the introduction of publication requirements emphasising Scopus-indexed journals led to an increase in the number of such articles published by Ukrainian scholars. However, to some extent, the quantity was prioritised over quality and Ukrainian academics also extensively published in local Scopus-indexed journals (Nazarovets, 2020; 2022; Hladchenko, 2022).

International mobility of Ukrainian scholars has been encouraged through promotion requirements for academic titles: to become an associate professor or professor, candidates must demonstrate a study visit or participation in a conference or symposium in OECD and/or European Union countries (Ministry of Education and Science of Ukraine, 2016). However, Ukrainian universities generally do not cover the costs of attending international conferences. Since the evening of February 24, 2022, Ukrainian male scientists between the ages of 18 and 60 have been subjected to restrictions on their right to freely travel outside of Ukraine. For them, international mobility is complicated by numerous legal and bureaucratic nuances, as permissions to travel abroad are limited and conditional (Chala et al., 2024).

**Academic migration from Ukraine**
Between 1994 and 1996, Ukraine's economic downturn triggered intensive academic migration, particularly among scholars in physics, mathematics, engineering, biology, and medical sciences, with many of the most talented researchers leaving the country (Ganguli, 2011). The rate of migration began to decline only after 2004 (Svityashchuk & Stadyi, 2014). From 1996 to 2009, 1,826 doctoral-level scholars emigrated, with the most popular destinations being the USA, Russia, Germany, Israel, and Canada (Kotlyar et al., 2009). Ukrainian scholars migrating to Germany were predominantly from cultural studies, philology, and engineering (Svityashchuk & Stadyi, 2014). Overall, the number of scholars in Ukraine fell dramatically from 313,079 in 1990 to 82,032 in 2012, a fourfold decrease (State Statistics Service of Ukraine, 2024). This decline resulted not only from migration but also from career changes to non-academic professions (Kupets, 2013). Scholars migrating abroad were typically aged 40 for PhD holders and 50 for those with second-level doctoral degrees. Kupets (2013) notes that by 2012, the financial benefits of migration had decreased two- to threefold compared to 2004–2005. Although opportunities for joint projects with foreign colleagues emerged, the persistent salary gap between scholars and other highly skilled professionals in Ukraine hindered the return of many migrants.

Svityashchuk and Stadyi (2014) argue that academic migration harms Ukraine, as it does any country, because highly capable individuals who could contribute to institutional development tend to leave countries with weak institutional capacity. Their departure reduces both human capital and the domestic demand for institutional improvements, with potentially severe long-term consequences.

**Data and methodology**
Since 2013, publications in Scopus-indexed journals have been increasingly incorporated into Ukrainian publication requirements. Accordingly, this study draws on Scopus data. The dataset was retrieved from the in-house version of Scopus maintained by the Centre for Science and Technology Studies (CWTS) at Leiden University.

The dataset was compiled in three steps. First, all scholars who had ever published with a Ukrainian affiliation were identified. Second, from this group, scholars who published with a foreign affiliation in 2020–2023 were selected. Third, only those whose prior publications included Ukrainian (co-)affiliations were retained. Two categories of scholars were excluded. The first comprised scholars affiliated with institutions located in Crimea, which was occupied by Russia in 2014. The second included scholars affiliated with institutions that Scopus listed as Russian in 2020–2023, although they are Ukrainian institutions in occupied territories. For these scholars, the change in affiliation reflects territorial occupation rather than individual mobility.

Ukrainian institutions were classified into three categories: universities, research institutes of the National Academy of Sciences of Ukraine (NASU), and other institutions. The final dataset includes 1,400 scholars: 618 who moved abroad in 2020–2021 and 782 in 2022–2023. Among them, 1,281 scholars had been (co-)affiliated with either universities or NASU prior to obtaining a foreign affiliation; these scholars constitute the core dataset for this study. The remaining 119 scholars were affiliated with other organisations, including hospitals (19), the National Academy of Agrarian Sciences (6), the National Academy of Educational Sciences (3), the National Academy of Medical Sciences (18), and the National Institute of Cancer (6).

Among the scholars in the core dataset, 29 were affiliated with both a university and NASU prior to moving abroad. Each of these scholars was counted twice—once in each category—resulting in 1,429 cases for calculating scholar numbers based on initial affiliation. Additionally, 35 scholars were affiliated with more than one country; these were treated as separate cases, yielding a total of 1,435 cases for analysis.

The research field of each scholar was determined based on the field in which they published the largest number of articles. Article research fields were classified according to the CWTS Scopus classification system, which includes biomedical & health sciences, life & earth sciences, mathematics & computer science, physical sciences & engineering, and social sciences & humanities. For 1,342 scholars, a single research area was assigned, while 58 scholars were associated with multiple research areas. In cases where scholars had an equal number of publications across multiple fields, their contributions were fractionalised among the respective fields.

Average field-normalised citation impact (FNCI) for each scholar was calculated based on articles published during the five years preceding their move abroad. Citations were normalised by publication year, research area, and publication type. In the core dataset, two scholars had no publications in the five years preceding their change of affiliation country.

Poisson linear regressions were used to statistically compare the average citation impact of scholars who moved abroad in 2020–2021 with those who moved in 2022–2023.

**Issues in scholars' affiliations in Scopus**

The initial dataset included a large number of internationally mobile Ukrainian scholars affiliated with non-EU countries, which prompted a verification of Scopus-listed affiliations against the affiliations stated in the original articles. Several types of mismatches were identified.

First, Scopus occasionally merges multiple scholars with similar last names and initials under a single AU-ID (e.g., AU-ID 56151580200, AU-ID 5623642790, AU-ID 57191967462, AU-ID 57193029606, AU-ID 571935419256, AU-ID 57193753323, AU-ID 57194002340, AU-ID 57218436437). Scholars may not request separation of these profiles, as a combined profile increases their publication and citation counts in Scopus.

Second, Scopus sometimes assigns affiliations that do not appear in the original articles. For instance, in article 'EID 84982918517', Scopus lists both Ukrainian and Russian affiliations for a scholar, whereas the original article includes only a Ukrainian affiliation. Similarly, Kharkiv Medical University appears as affiliated with Russia in articles 'EID 85182680531' and 'EID 85182657711', despite never being under Russian occupation. Bogomolets Kyiv Medical University is also misattributed in several articles ('EID 85176388753', repeated). Several affiliations from the *Eastern-European Journal of Enterprise Technologies*, a Ukrainian journal, are incorrectly recorded in Scopus. In some cases, when Scopus cannot extract the exact Ukrainian affiliation, it assigns the scholar to another non-EU country; for example, some scholars were incorrectly affiliated with Mozambique and Vietnam.

Third, mismatches can result from ambiguous or improperly stated affiliations in the original articles. For example, article 'EID 201180155174981' lists three affiliations without specifying the corresponding authors. Scopus then assigns all three affiliations to all authors. In cases where a scholar's affiliation country is missing, Scopus may default to the affiliation of the following author.

**Results**

**Mobility patterns across NASU and universities**

Figure 1 illustrates the affiliation distribution of Ukrainian scholars from 2020 to 2023. It shows an increase in the share of internationally mobile scholars as well as those co-affiliated with foreign organisations for both NASU and universities. In 2023, the proportion of internationally mobile scholars reached 1.9% for the NASU and 2.9% for universities. While universities showed an increase in internationally mobile scholars, the NASU experienced a more pronounced rise in scholars with foreign co-affiliations.

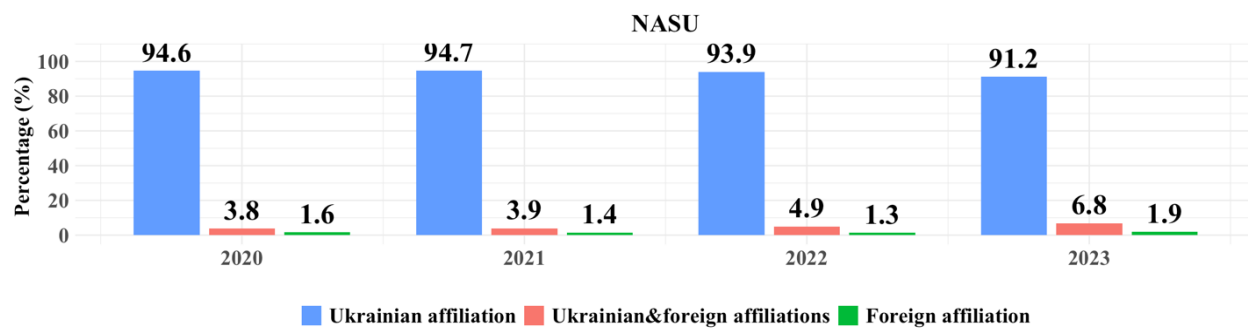

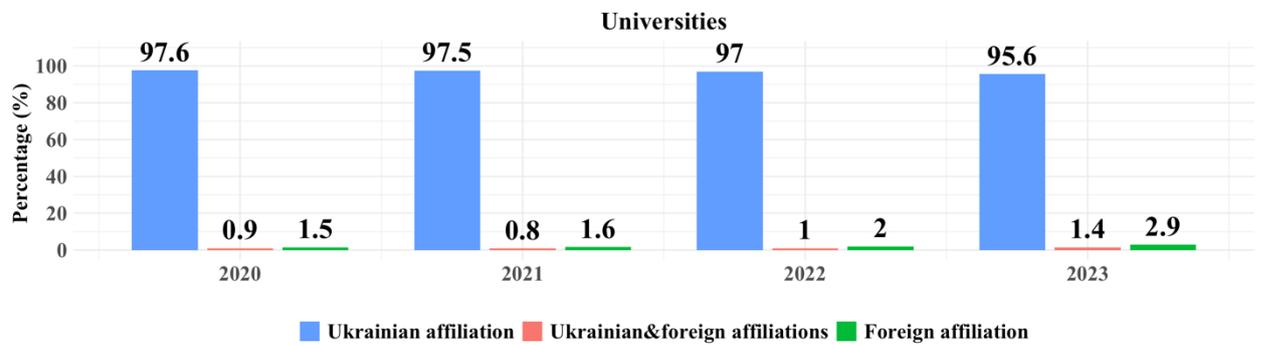

**Fig. 1** Affiliation distribution of Ukrainian scholars

Figure 2 shows that the number of internationally mobile Ukrainian scholars grew from 618 in 2020–2021 to 782 in 2022–2023. This increase was driven primarily by scholars previously affiliated with universities, whose number rose from 280 (44.4% of the total) to 437 (54.7%). By contrast, both the number and share of scholars with dual Ukrainian and foreign affiliations declined. For scholars affiliated with the National Academy of Sciences of Ukraine (NASU), the absolute number increased slightly from 120 to 136, although their share decreased from 19.0% to 17.1%.

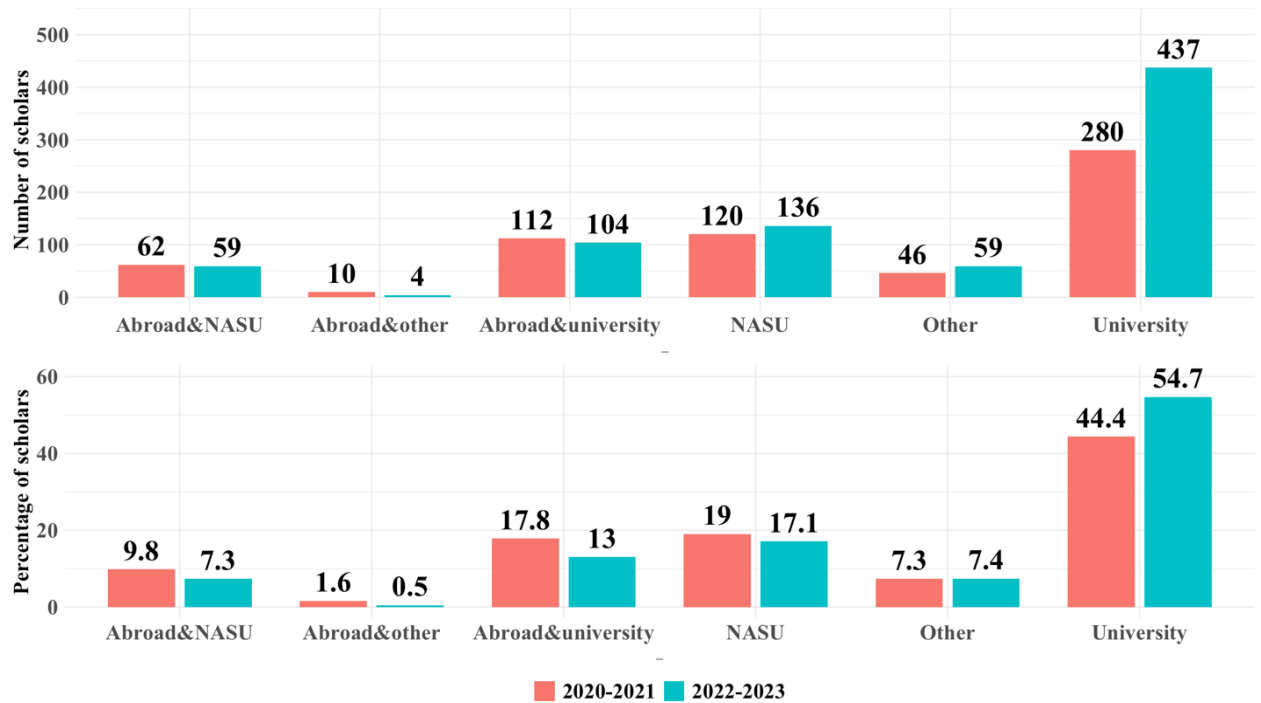

**Fig. 2** Distribution of internationally mobile scholars by initial affiliation

### Distribution of internationally mobile scholars across research fields

Figure 3 illustrates the distribution of internationally mobile scholars by research field. For the NASU, the physical sciences & engineering dominated in both periods. During 2022–2023, their share increased, alongside a rise in the social sciences, while other research fields experienced a decline. In both periods, biomedical & health sciences were the second in size.

For universities, physical sciences & engineering accounted for more than half of internationally mobile scholars in 2020–2021. However, their share fell to 39.3% in 2022–2023, reflecting a relative rise in other research fields. The social sciences & humanities constituted the second largest field, increasing from 17% to 20%. This pattern suggests that the war-induced displacement prompted a broader diversification of internationally mobile scholars across research fields within universities.

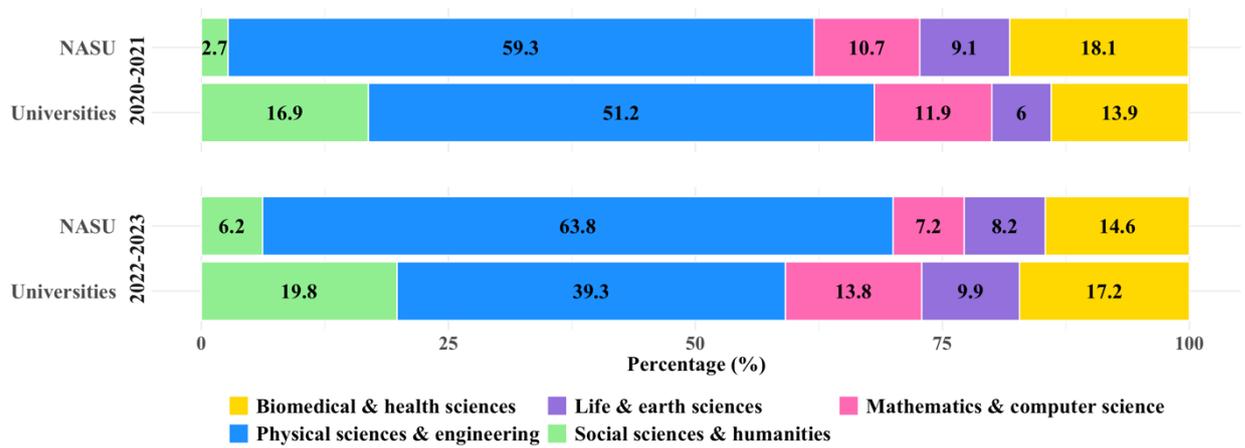

**Fig. 3** Internationally mobile scholars by research field

**Top ten destination countries**

The top ten destination countries for the NASU and universities are presented in Figures 4-7. For the NASU, Russia (18,4%, N=34) was the top host country during 2020-2021, followed by Poland (11,9%, N=22) and Germany (11,4%, N=21). In 2022–2023, Germany (14.0%, $N = 28$), China (11.5%, $N = 23$), and Poland (10.5%, $N = 21$) emerged as the leading host countries, indicating an increase in the number of Ukrainian internationally mobile scholars. In contrast, Russia (9.5%, $N = 19$) declined to fourth position. Although the absolute number of scholars moving to Poland remained almost unchanged, its share of total mobility decreased from 11.9% to 10.5%, moving it from second to third place. Several new destination countries also entered the top ten in 2022–2023, including Israel and Austria, reflecting a diversification of host countries for Ukrainian scholars. In 2022-2023, Austria, Czechia, France, Italy and Sweden each hosted the same number of Ukrainian internationally mobile scholars.

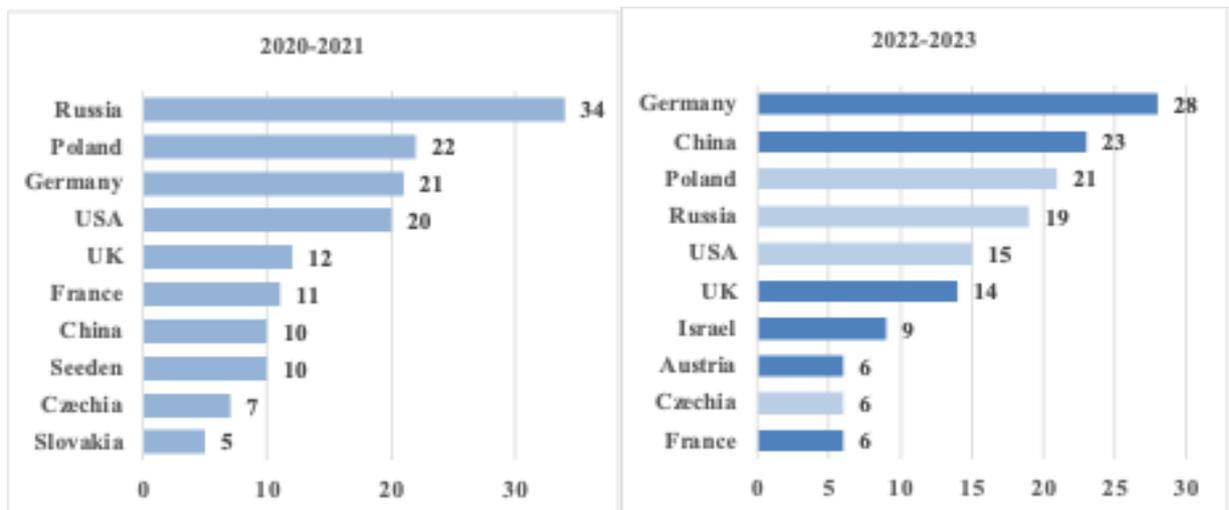

**Fig. 4** Top ten destination countries for scholars affiliated with NASU (number)

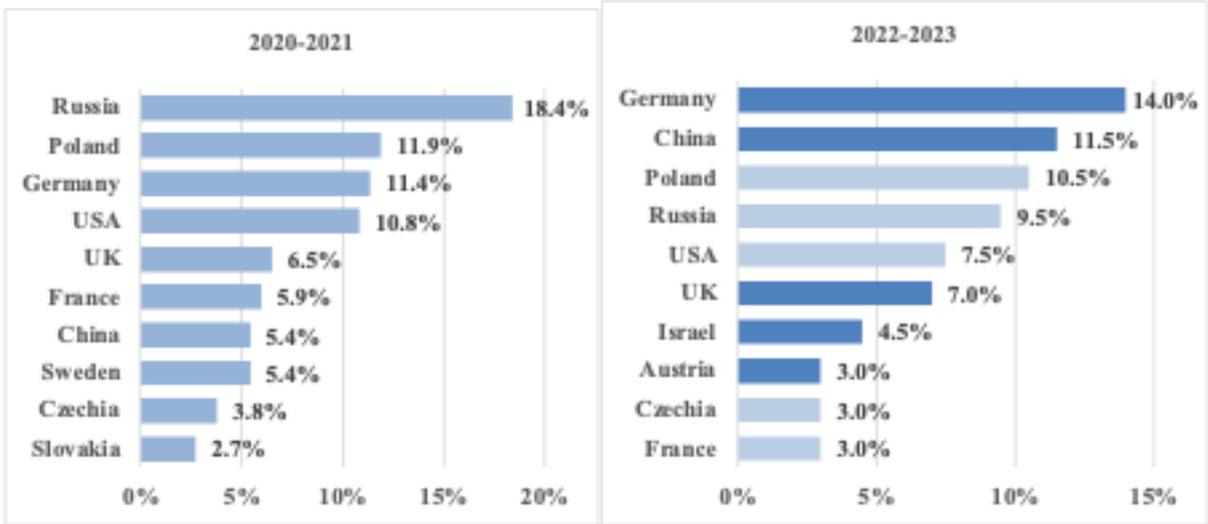

**Fig. 5** Top ten destination countries for scholars affiliated with NASU (percentage)

Figures 6 and 7 demonstrate that, for universities in both periods, the largest numbers of scholars moved to Poland, Germany, Russia, and the United States. In 2022–2023, the absolute number of scholars who moved to the top ten destination countries increased, although the corresponding percentages did not rise uniformly. Mobility to neighbouring Eastern European countries—such as Poland, Czechia, Slovakia, and Lithuania—was more pronounced among university scholars than among those from NASU and showed further growth during 2022–2023.

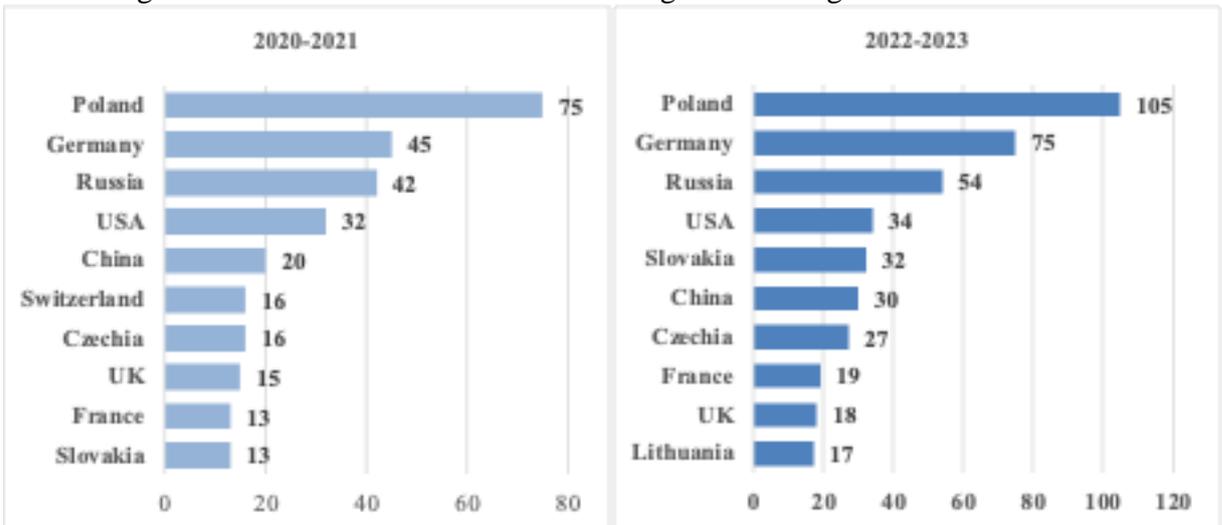

**Fig. 6** Top ten destination countries for scholars affiliated with universities (number)

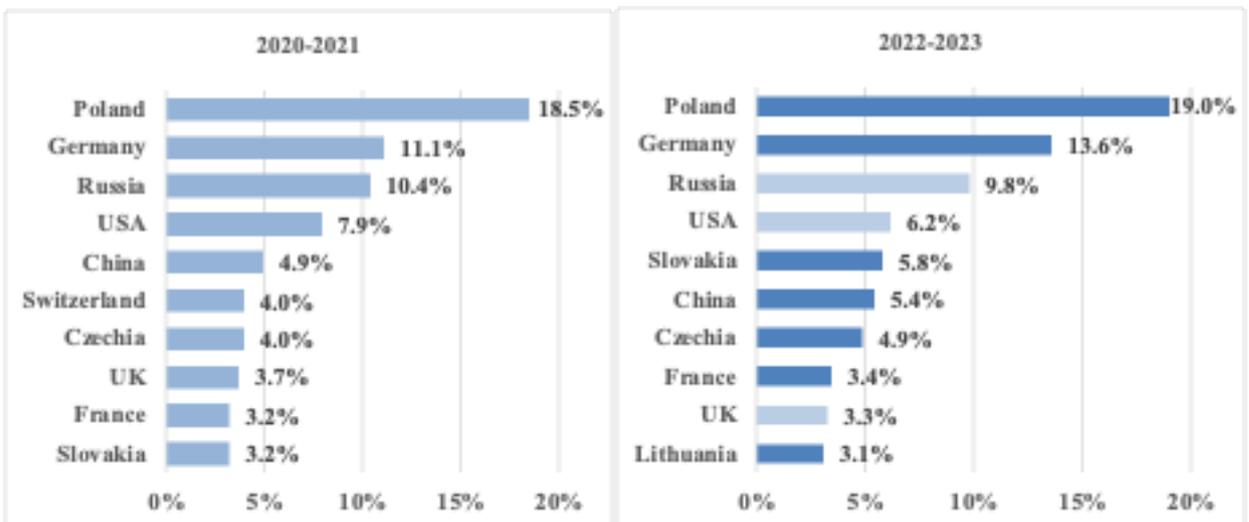

**Fig. 7** Top ten destination countries for scholars affiliated with universities (percentage)

**Top ten destination countries and citation impact**

Figures 8 and 9 show the distribution of average FNCI per scholar across the top ten destination countries for scholars previously affiliated with NASU. In 2020–2021, the highest mean FNCI was observed among scholars who moved to the UK (0.79), while those who moved to Russia and Poland both recorded 0.64, and Germany 0.52. In 2022–2023, the UK again ranked first, with a substantially higher mean FNCI of 1.11. While in 2020-2021, highly cited scholars were relatively evenly distributed among most countries, in 2022-2023, they became concentrated in Germany, the USA, and the UK. Germany's mean FNCI remained relatively stable (0.48 vs. 0.52), as did that of China (0.47 vs. 0.45), whereas Russia experienced a notable decline (0.60 vs 0.19).

For universities, scholars who moved to Germany (0.71) and Switzerland (0.66) had the highest mean FNCI in 2020-2021 (Fig. 10). In 2022-2023 the highest mean FNCI had scholars who moved to the UK (1.55), Slovakia (0.97) and Poland (0.88) (Fig. 11). High mean FNCI for these countries were due to highly cited scholars. Apart from Germany, the mean FNCI for most countries rose. The number of highly cited scholars also increased, as well as there were higher average FNCI per scholar compared to 2020-2021. The largest number of highly cited scholars was among those who moved to Poland.

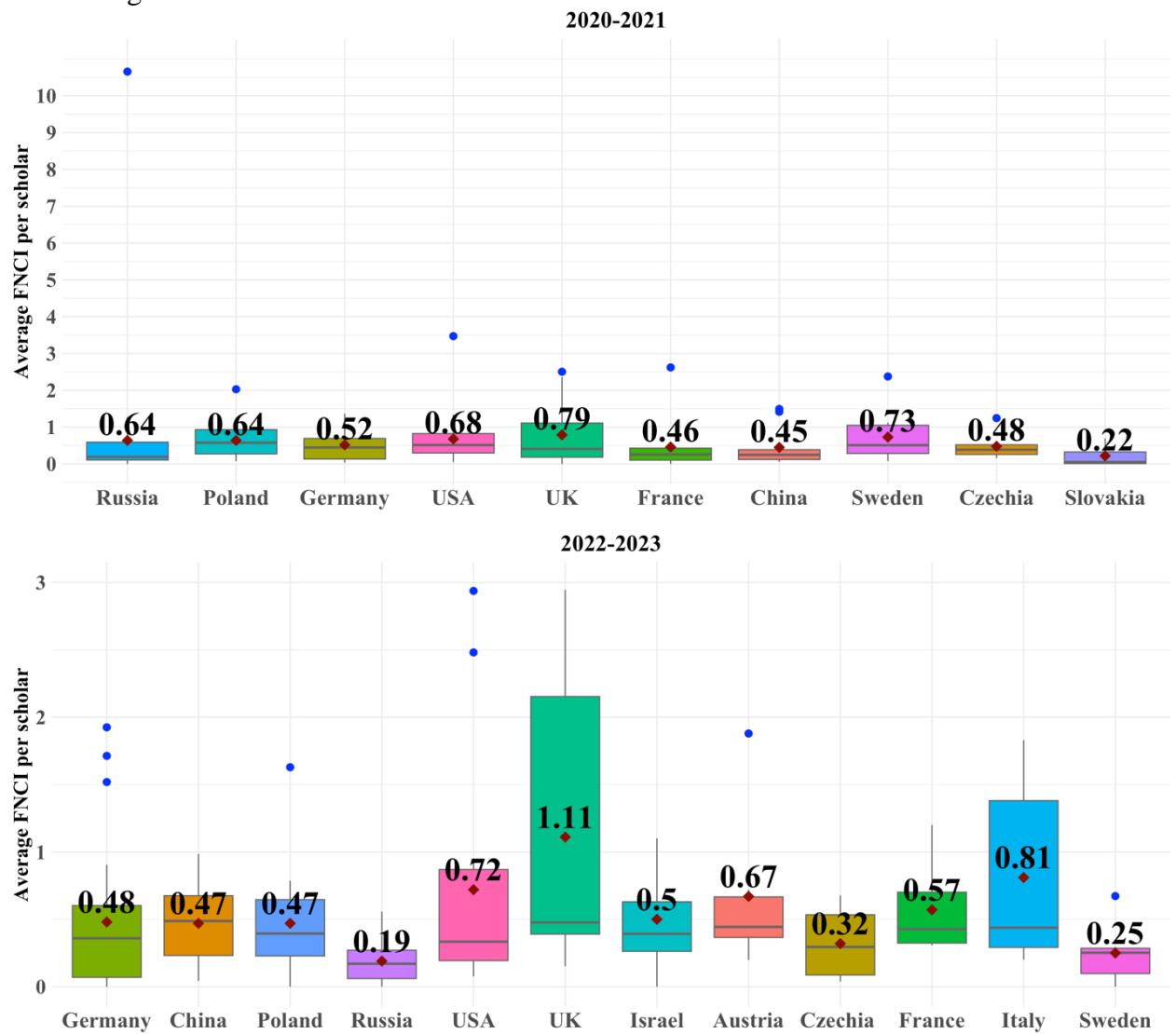

**Fig. 8** Average FNCI per scholar across the top ten destination countries (NASU)

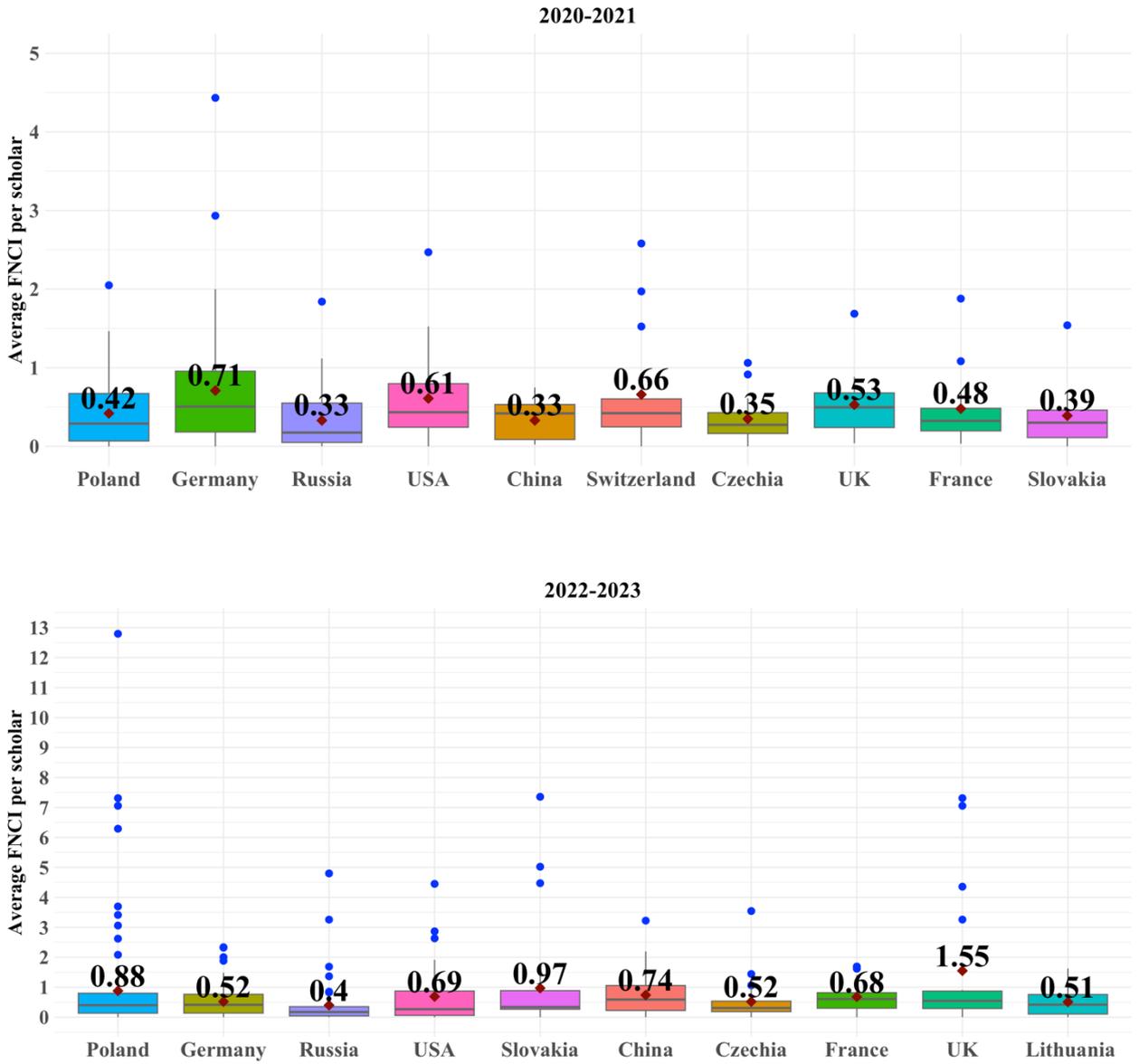

**Fig. 9** Average FNCI per scholar across the top ten destination countries (universities, 2022-2023)

.

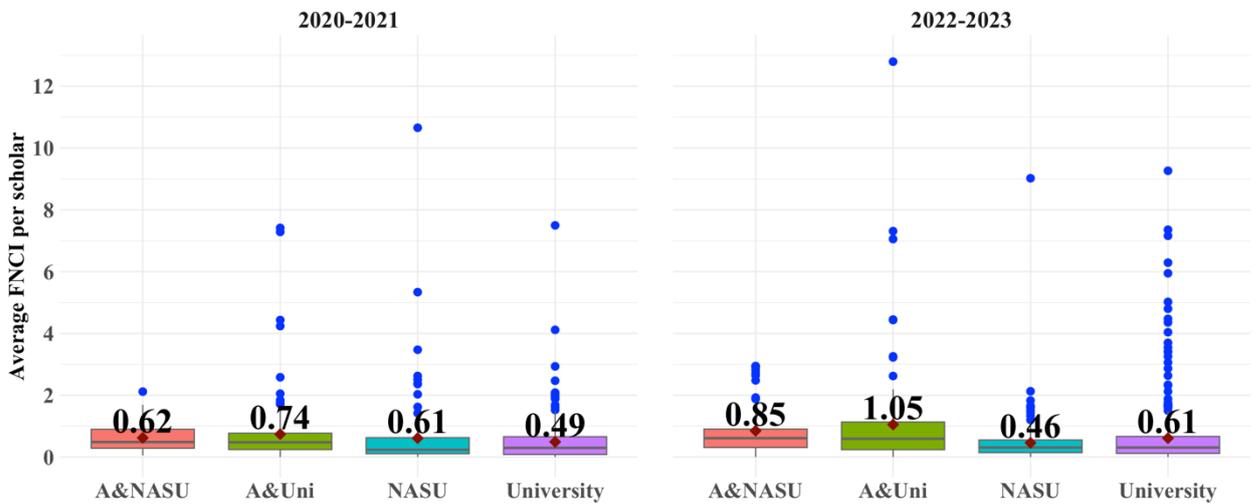

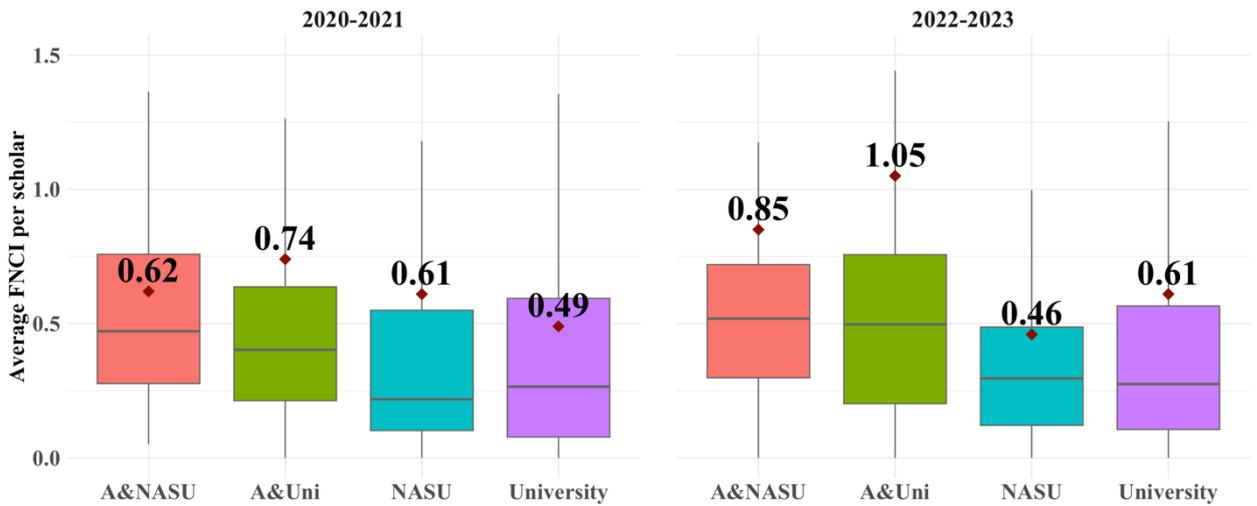

**Fig. 10** Average FNCI per scholar across initial affiliations

Figure 10 illustrates a comparison of average FNCI per scholar across initial affiliations. It reveals a rise in mean FNCI across all initial affiliations except for NASU. In both periods, it was the highest for scholars previously affiliated with both a university and a foreign institution. The number of highly cited scholars increased substantially among scholars previously affiliated with universities.

The Poisson regression was used to compare the average FNCI of scholars between two periods (2020–2021 and 2022–2023) while considering different types of affiliations (Table 2). The results indicate that there was no significant difference in average FNCI between the two periods overall (Model 1), nor across affiliations in 2020–2021 (Model 2). In contrast, in 2022–2023 (Model 3), scholars affiliated with NASU ($\beta$ = –0.609, p = 0.001) and universities ($\beta$ = –0.337, p = 0.028) displayed lower FNCI, while no systematic period effect was observed for the Abroad & NASU and Abroad & university subgroups (Models 4–5). At the same time, university-affiliated scholars and those affiliated with both university and foreign institution (Models 6–7) showed small positive effects in 2022–2023 ($\beta$ = 0.349, p = 0.016; $\beta$ = 0.217, p = 0.039). However, across all Poisson models, the pseudo-$R^2$ values remained very low (0.003–0.048), indicating that these findings likely reflect the influence of small subgroups rather than broad systematic changes in FNCI across the population.

**Table 2** Results of Poisson regression models comparing FNCI across affiliations and periods

|  | Model1 2020-2021 vs 2022-2023 | Model 2 2020-2021 | Model 3 2022-2023 | Model 4 Abroad & NASU | Model 5 NASU | Model 6 Abroad & university | Model7 University |
|---|---|---|---|---|---|---|---|
| Intercept | -0.548*** (<2e-16) | -0.480** (0.003) | -0.160(0.256) | -0.480*** (0.0001) | -0.491*** (0.002) | -0.299** (0.006) | -0.714*** (5.81e-12) |
| 2022-2023 | 0.137 (0.133) |  |  | 0.320(0.136) | -0.278 (0.106) | 0.349* (0.016) | 0.217* (0.039) |
| Abroad & university |  | 0.180(0.356) | 0.210(0.218) |  |  |  |  |
| NASU |  | -0.011(0.955) | -0.609**(0.001) |  |  |  |  |
| University |  | -0.235(0.199) | -0.337*(0.028) |  |  |  |  |
| Pseudo-R² | 0.0033 | 0.0215 | 0.0478 | 0.0382 | 0.0124 | 0.0226 | 0.0075 |
| Obs. | 1308 | 572 | 736 | 121 | 256 | 216 | 715 |

Significance levels:***p< 0.001,**p < 0.01, *p < 0.05

**Scholars with articles among the top 10% most cited globally**

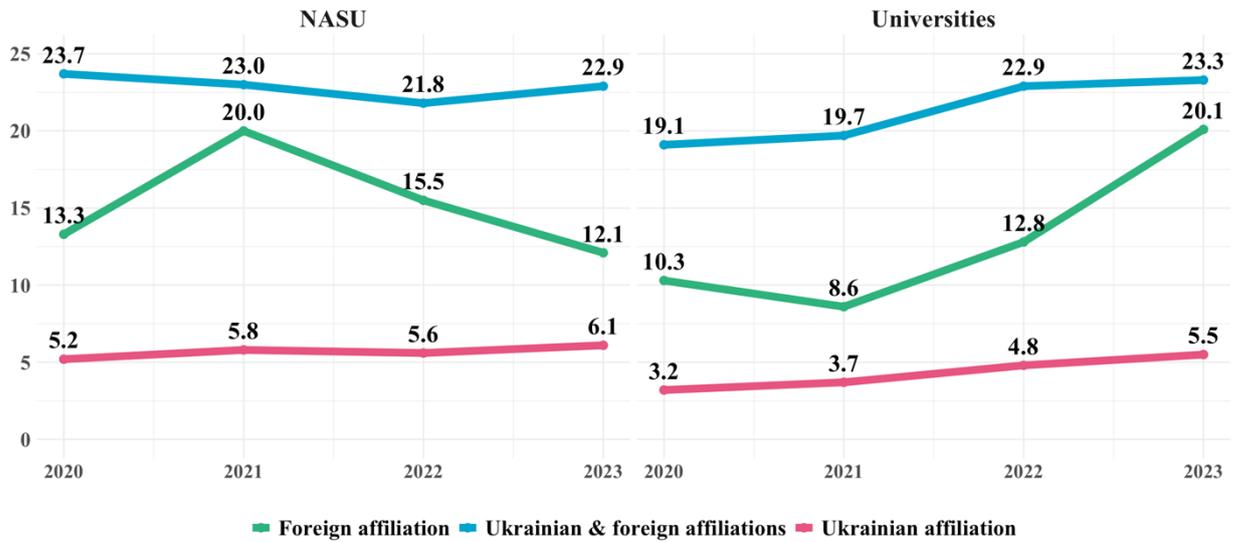

**Fig. 11** The percentage of scholars with articles among the top 10% most cited globally across affiliation patterns

Figure 11 shows that, from 2020 to 2023, for both NASU and universities, the percentage of scholars with articles among the top 10% most cited globally was the highest among those affiliated with both Ukrainian and foreign institutions. Conversely, it was the lowest among scholars affiliated solely with Ukrainian institutions. Among scholars who moved abroad, this percentage decreased for NASU from 13.3 to 12.1% and increased for universities from 10.3 to 20.1%. For both NASU and universities, the percentage of scholars with articles among the top 10% most-cited globally is higher for those who moved abroad than for those remaining in Ukraine, suggesting that the most productive scholars tend to leave the country. On the positive side, the number of top-performing scholars affiliated solely with Ukrainian institutions slightly increased for both NASU and universities.

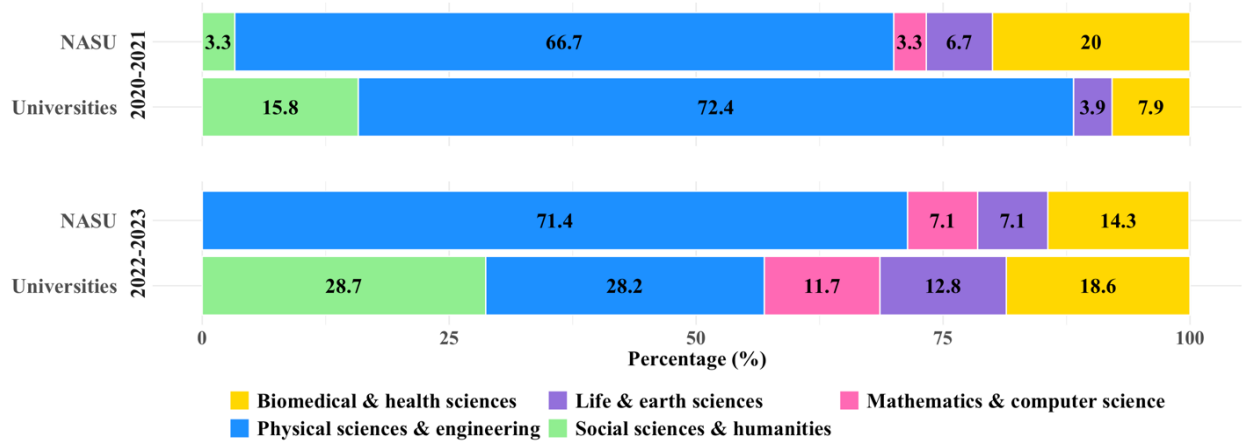

**Fig. 12** The percentage of internationally mobile scholars with articles among the top 10% most cited globally by research field

Figure 12 illustrates the distribution of scholars with articles among the top 10% most cited globally across research fields. For NASU, scholars in physical sciences & engineering constituted the largest share in both periods, increasing from 64% to 71%. Although the share of scholars in social sciences & humanities grew slightly between the two periods in the total research output of NASU, they were not represented among those with articles in the top 10% most-cited globally.

For universities, the share of scholars in physical sciences & engineering decreased from 72.4% to 28.2%, reflecting a relative increase in scholars from other disciplines and the emergence of scholars from mathematics & computer science in this category. The share of scholars from social sciences & humanities increased from 15.8% to 28.7%, becoming the largest among university-affiliated scholars.

Figure 15 presents the percentage of scholars with articles among the top 10% most cited globally across the top ten destination countries for NASU and universities (i.e., as a share of the total number of NASU scholars and the total number of university scholars, respectively). For NASU, Russia had the highest percentage in 2020–2021, followed by Poland and the UK. In 2022–2023, Germany and the UK took the lead. The percentage of top-performing scholars increased among those affiliated with Germany and the UK but decreased among those affiliated with Poland. Notably, during this later period, no scholars with articles in the top 10% most cited globally moved to Russia or China.

For universities, in 2020-2021, the scholars with articles among the top 10% most cited globally were only in five destination countries (Poland, Germany, Russia, the USA and China), in 2022-2023, the range of countries increased, with Poland and Germany accumulating the highest percentage in both periods.

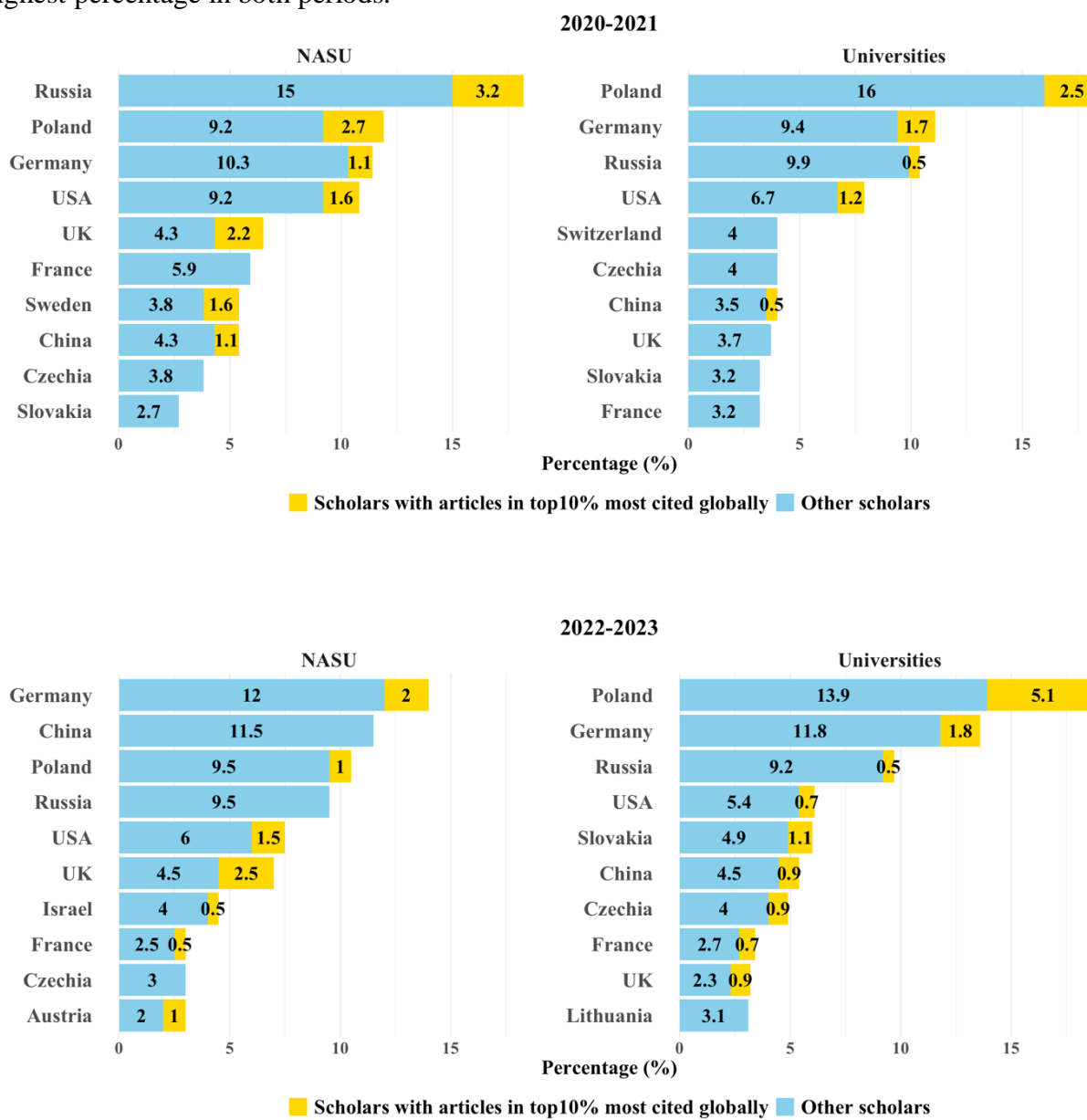

**Fig. 13** The percentage of scholars with articles among the top 10% most cited globally across the top ten destination countries (2022-2023)

Figure 14 presents the top five destination countries for scholars with articles among the global top 10% most cited, calculated separately as a share of the total number of top-cited NASU scholars and top-cited university scholars. For NASU, they included Russia, Poland, and the UK during 2020-2021, and the UK, Germany, and Italy during the following period. For universities, Poland, Germany and the USA took the lead during 2020-2021, and Poland, Germany and Slovakia during 2022-2023. For universities, Poland and Germany, that accumulated the largest number of internationally mobile Ukrainian scholars also accumulated the largest number of scholars with articles among the top 10% most cited globally.

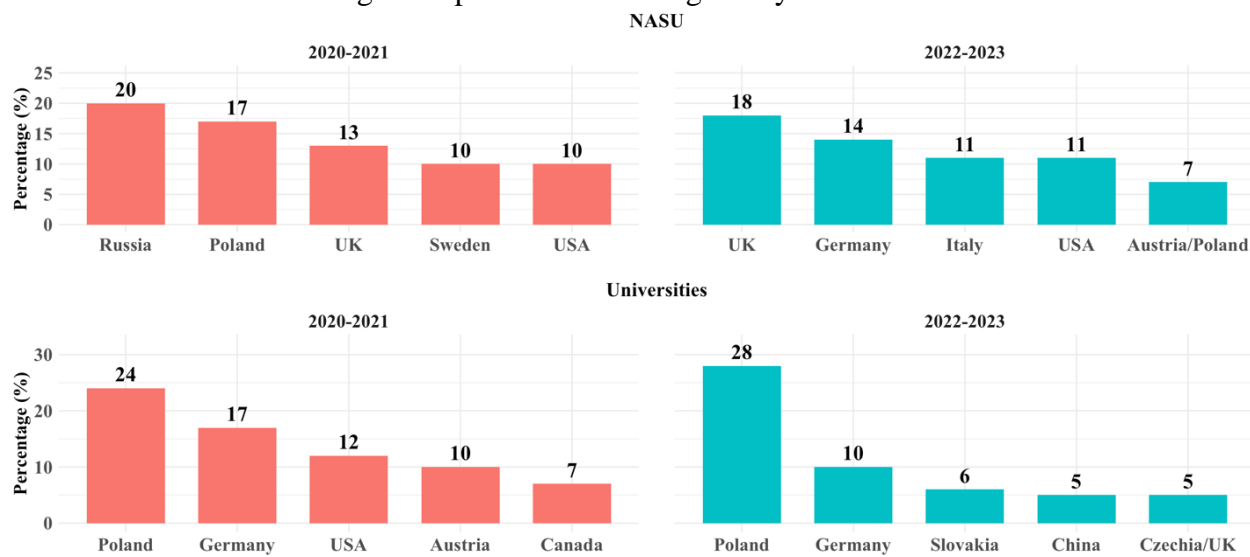

**Fig. 14** Top five destination countries for scholars with articles among the top10% most cited globally

**Conclusions**

*Issues in scholars' affiliations in Scopus*

Issues faced during the checking of the accuracy of scholars' affiliations stress the necessity for journals to adopt a standardised layout for author data, including affiliations. If Scopus and other indexing services required such a format, the accuracy of bibliometric data would be significantly improved. As bibliometric data increasingly inform national and institutional research assessments, their accuracy is crucial for data-driven policymaking.

*Trends in the number and disciplinary distribution of internationally mobile Ukrainian scholars*

The study findings revealed an increase in the number of internationally mobile Ukrainian scholars during the first two years of Russia's full-scale invasion. This rise was mainly observed among scholars previously affiliated with universities. For NASU, the growth was less pronounced among internationally mobile scholars but more substantial among those holding foreign co-affiliations. Since NASU positions do not include teaching, it may have been legally easier for these scholars to maintain dual affiliations**.**

   In terms of disciplines, scholars from physical sciences & engineering constituted the major share of internationally mobile scholars irrespective of their initial affiliation. This finding aligns with prior studies showing that physical sciences & engineering exhibit the highest levels of geographical mobility (Petersen, 2018; Hunter et al., 2009; Sincell, 2000; Stephan, 2012; Scellato et al., 2012). However, while the share of internationally mobile scholars from physical sciences & engineering increased among NASU scholars, it declined for universities, where the distribution of disciplines became more balanced. This pattern can be partly attributed to disciplinary distinctions between NASU and universities. It may also be influenced by constraints on male scientific mobility under martial law, given that male scholars are predominant in STEM disciplines. A limitation of this study is that it did not analyse international mobility from a gender perspective, nor did it consider the potential impact of martial law on male scholars' mobility.

*Main destination countries*

In both periods, the top destination countries included neighbouring countries (Russia, Poland, Czechia, Slovakia) as well as global scientific hubs (Germany, China, the USA, and the UK). Mobility to neighbouring Eastern European countries—such as Poland, Czechia, Slovakia, and Lithuania—was more pronounced among university scholars than NASU scholars, and its volume increased during 2022–2023. This aligns with prior research highlighting the role of geographic and cultural proximity (Mahroum, 2000; Kiselyova & Ivashchenko, 2024; El-Ouahi et al., 2021). For universities, mobility to neighboring countries increased not only due to proximity but also because of substantial financial initiatives introduced after 2022. During 2022–2023, NASU scholars shifted their mobility focus from Russia and Poland toward Germany and China, with Poland ranking third.

Germany, China, the USA, and the UK, which were among the top destination countries for both NASU and university scholars, represent so-called global hubs for scientific mobility (Grogger & Hanson, 2011; Aref et al., 2019; Van Noorden, 2012; Franzoni, Scellato & Stephan, 2012; Ganguli, 2015). The rise in scholars moving to Germany during 2022–2023 reflects the country's status as a leading destination for displaced scholars in Europe, partly due to abundant third-party funding opportunities (Vatansever, 2022). The growing volume of scholars migrating to the UK and the USA supports prior findings that scholars prioritise English-speaking countries (Adsera & Pytlikova, 2015), which aligns with English being the most widely taught foreign language in Ukraine, as well as financial initiatives provided by the UK to host Ukrainian scholars.

The top ten destination countries and countries of international collaboration partly, but not entirely, overlap for both NASU and universities. Notably, for universities, Poland was also a top country of international collaboration during both periods (Hladchenko, 2025). The partial overlap between these two groups of countries suggests that physical relocation and research collaboration follow related but distinct patterns and depend on corresponding financial initiatives.

Importantly, the most highly cited scholars exhibited distinctive mobility patterns: for NASU, the UK and Germany overtook Russia and Poland as primary destinations in 2022–2023, whereas for universities, Poland and Germany remained the main hubs.

*Citation impact*

A statistical test comparing the average FNCI per scholar for those who moved in 2020–2021 and 2022–2023 revealed no significant overall differences. However, visualisation and subgroup analysis showed modest improvements for university-affiliated scholars, driven by a rise in the number of highly cited individuals among those who moved abroad. Similarly, the share of scholars with articles in the global top 10% most cited increased for universities, but declined for NASU.

*Concluding reflections and policy directions*

To summarise, Russia's full-scale invasion has triggered both an increase and diversification in Ukrainian scholars' international mobility. While the rise in mobility risks deepening brain drain—since highly cited scholars are disproportionately represented among those leaving—the situation also creates opportunities for "brain circulation."

Currently, internationally mobile Ukrainian scholars are enhancing their professional skills, accessing advanced resources, and building new networks. For Ukraine to benefit from these dynamics, policymakers and universities must adopt strategies that transform forced mobility into long-term gains, encouraging return, engaging the diaspora, and supporting dual affiliations. Similar policies in China have successfully transformed brain drain into brain circulation (Cao et al., 2020; Marini & Yang, 2021). If harnessed effectively, forced international mobility could help integrate Ukraine more deeply into global scientific networks, accelerate post-war recovery, and reshape its academic system into one that is more resilient, internationally connected, and competitive.

**Acknowledgement**

The author is grateful to Rodrigo Costas Comesana for comments and suggestions to this article.

**Declarations**
**Conflict of interest** The author claims no conflict of interest